# Tutorial on logistic-regression calibration and fusion: Converting a score to a likelihood ratio

Geoffrey Stewart Morrison[*]

*Forensic Voice Comparison Laboratory, School of Electrical Engineering & Telecommunications, University of New South Wales, Sydney, Australia*

Logistic-regression calibration and fusion are potential steps in the calculation of forensic likelihood ratios. The present paper provides a tutorial on logistic-regression calibration and fusion at a practical conceptual level with minimal mathematical complexity. A score is log-likelihood-ratio like in that it indicates the degree of similarity of a pair of samples while taking into consideration their typicality with respect to a model of the relevant population. A higher-valued score provides more support for the same-origin hypothesis over the different-origin hypothesis than does a lower-valued score; however, the absolute values of scores are not interpretable as log likelihood ratios. Logistic-regression calibration is a procedure for converting scores to log likelihood ratios, and logistic-regression fusion is a procedure for converting parallel sets of scores from multiple forensic-comparison systems to log likelihood ratios. Logistic-regression calibration and fusion were developed for automatic speaker recognition and are popular in forensic voice comparison. They can also be applied in other branches of forensic science, a fingerprint/finger-mark example is provided.

This research was supported by the Australian Research Council, the Australian Federal Police, New South Wales Police, Queensland Police, the National Institute of Forensic Science, the Australasian Speech Science and Technology Association, and the Guardia Civil via Linkage Project LP100200142. Thanks to Julien Epps, Max Welling, Niko Brümmer, David Balding, Ian Evett, Cedric Neumann, and three anonymous reviewers for comments on earlier versions of this paper. Unless otherwise explicitly attributed, the opinions expressed herein are those of the author and do not necessarily represent the policies or opinions of any of the above mentioned organisations or individuals.

## Introduction

This tutorial paper describes logistic-regression calibration and fusion, each of which is a potential step in the calculation of likelihood ratios. The procedures were developed for automatic speaker recognition and have become popular in forensic voice comparison, they have been applied in numerous papers in this branch of forensic science, including references 1, 2, 3, 4, 5, 6, 7, 8, 9, 10, 11, 12, 13, 14, 15. They are also potentially applicable in other branches of forensic science. Indeed the importance of calibration was raised by several of the discussants of a recent paper on calculating likelihood ratios for fingerprint/finger-mark comparisons (Neumann et al[16]), and calibration was discussed in at least four presentations in the "strength of evidence" theme at the recent 2012 European Academy of Forensic Science Conference[17,18,19,20]. Logistic-regression calibration is also common practice in machine learning, with a seminal paper

---

[*]e-mail: geoff-morrison@forensic-voice-comparison.net

in that field being Platt[21].

It is common for the early stages of forensic-voice-comparison systems to produce scores; a score is log-likelihood-ratio like in that it indicates the degree of similarity of a pair of samples while taking into consideration their typicality with respect to a model of the relevant population (the need in the current context for a score to take account of typicality as well as similarity is discussed in the "Application to scores generated in other ways" section below). A higher-valued score provides more support for the same-origin hypothesis over the different-origin hypothesis than does a lower-valued score; however, the absolute values of scores are in general not interpretable as log likelihood ratios. Logistic-regression calibration is a procedure for converting scores to log likelihood ratios, and logistic-regression fusion is a procedure for converting parallel sets of scores from multiple forensic-comparison systems to log likelihood ratios. (By "forensic-comparison system" I mean a set of procedures and databases which are used to compare two samples, one of known origin and one of questioned origin, and produce a likelihood ratio. Systems are broadly defined and could be fully automatic or could include elements of human supervision, they could be based on measurements and statistical models or could be based on an expert's experience-based opinion.)

I have taught logistic-regression calibration and fusion on multiple occasions as part of graduate-level courses and as part of tutorials associated with conferences, but have not found a written description of the procedures available at an appropriate introductory level, or from the perspective of calculating likelihood ratios interpretable as estimates of forensic strength of evidence. The present paper therefore describes the procedure from a forensic-likelihood-ratio-calculation perspective and at a practical conceptual level with minimal mathematical complexity. More in-depth discussion of fundamental theory and technical detail can be found in the automatic-speaker-recognition papers listed in references 22, 23, 24, 25, 26.

For concreteness, I will tend to describe the procedures as applied in forensic voice comparison, but the presentation should be accessible to readers without a background in this particular branch of forensic science, and I provide one example applied to scores generated from fingerprint/finger-mark data. The reader is assumed to be familiar with the likelihood-ratio framework at a philosophical/theoretical level and to understand the basics of how to calculate a likelihood ratio.

I first describe a simple procedure for calculating likelihood-ratios using single Gaussians to model the suspect and the population and using a single data point from the offender sample. I then introduce a problem which arises when, as is typical in forensic voice comparison, there are multiple data points from the offender sample. I present a solution to this problem based on calculating scores and then converting those scores to likelihood ratios. A procedure for converting scores to likelihood ratios using single Gaussians is described, it is shown that under certain assumptions this is equivalent to logistic-regression calibration, but that logistic regression is more robust. Applications of logistic-regression calibration to scores generated in other ways, including human-generated similarity judgements, are discussed. Although this paper introduces calibration as a solution to the multiple-data-point problem, it also discusses calibration as good practice to ameliorate results which are miscalibrated due to other causes. Finally, logistic-regression fusion is briefly described, and examples of both calibration and fusion are presented.

**A simple procedure using single Gaussians to calculate a likelihood ratio**

I begin by describing a simple procedure for calculating a likelihood ratio. Imagine we have univariate (one-dimensional) continuously-valued data, say the average fundamental frequency measured over a recording of a speaker (for generality we will represent such a measurement by the symbol $x$). We have a large number of data points from a large number of speakers representative of the relevant population (a background sample), multiple data points from a suspect, and a single data point from the offender. We build a single-Gaussian model of the background (representing the different-origin hypothesis) and a single-Gaussian model of the suspect (representing the same-origin hypothesis), and then evaluate the probability-density-function (pdf) value of each model at the value of the data-point from the offender, see Figure 1 and Equation 1. The pdf value from the suspect model relative to the pdf value from the background model (the relative heights of the model curves in Figure 1) is our calculated likelihood ratio.

$$LR = \frac{f(x_{off} | H_{so})}{f(x_{off} | H_{do})} = \frac{\frac{1}{\sigma_{susp}\sqrt{2\pi}} e^{\frac{(x_{off} - \mu_{susp})^2}{-2\sigma_{susp}^2}}}{\frac{1}{\sigma_{back}\sqrt{2\pi}} e^{\frac{(x_{off} - \mu_{back})^2}{-2\sigma_{back}^2}}} \qquad (1)$$

Where $LR$ is the likelihood ratio; $f(x_{off} | H_{so})$ and $f(x_{off} | H_{do})$ are the probability-density-function values, from same-origin and different-origin models respectively, evaluated at $x_{off}$, the average fundamental-frequency value extracted from the offender recording; $\mu_{susp}$ and $\mu_{back}$ are the means of the average fundamental-frequency values calculated from the suspect and background recordings respectively; and $\sigma_{susp}$ and $\sigma_{back}$ are the standard deviations of the average fundamental-frequency values calculated from the suspect and background recordings respectively. Data are invented for illustrative purposes and are not meant to be realistic representations of fundamental-frequency measurements of any particular speaker or group of speakers.

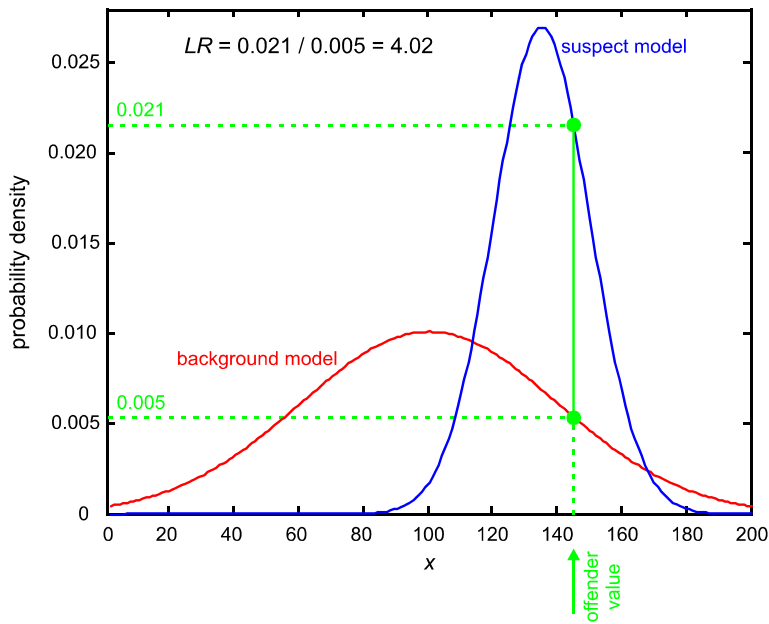

Figure 1: Example of the calculation of a likelihood-ratio value at a given raw-data value for the offender using single-Gaussian suspect and background models. Data are invented for illustrative purposes.

**More complex models**

The single-Gaussian procedure can trivially be expanded to multivariate data using mean vectors and covariance matrices rather than scalar means and variances. The approach can also be generalised to use any probability density function, and using a more complex model would be appropriate in cases where it would be inappropriate to assume that the underlying data have a single-Gaussian distribution. Appropriate models could be arbitrarily complex and the Gaussian-mixture model - universal background model approach (GMM-UBM, e.g., Reynolds, Quatieri, and Dunn[28]) is common in forensic voice comparison. Figure 2 illustrates the calculation of a likelihood ratio on the basis of more complex pdf models (the example is univariate but the procedure is also applicable to multivariate data).

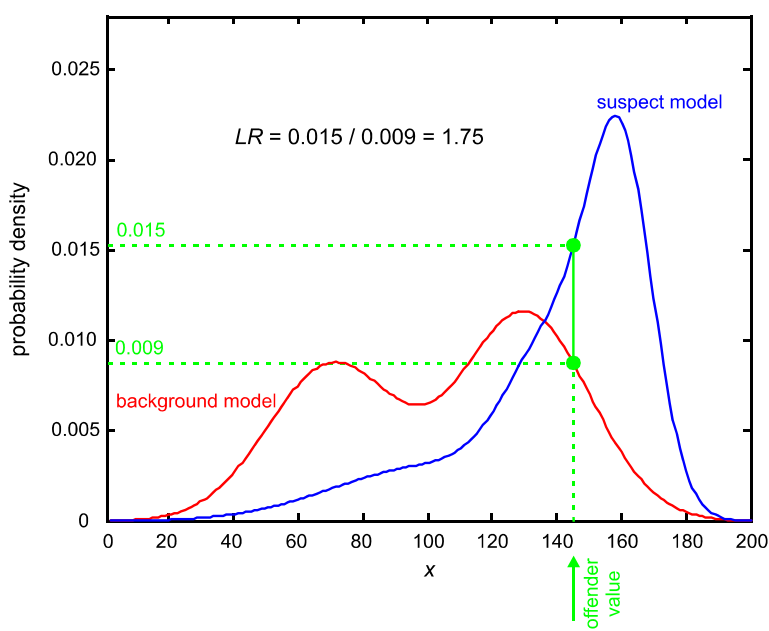

Figure 2: Example of the calculation of a likelihood-ratio value at a given raw-data value for the offender using Gaussian-mixture suspect and background models. Data are invented for illustrative purposes.

**A problem with multiple data points from the offender sample**

A problem arises if we have more than one data point from the offender sample, and in fact this is almost always the case in forensic voice comparison. Examples of two common scenarios are as follows: In one scenario numeric data are extracted from the voice recording every few milliseconds, mel-frequency cepstral coefficients (MFCCs) are typically extracted by first placing a 20-ms-wide window on the signal, performing calculations on that portion of the signal, then advancing the window by 10 ms, and repeating the steps above until the whole speech-active portion of the recording has been analysed. This results in a large number of data points which characterise the spectral properties of the signal. The other scenario is that multiple tokens of a particular phonetic unit are selected in the recording and each of these units is acoustically analysed, each producing a data point (phonetic units could be phoneme, word, or phrase length, e.g., all the tokens of "like" in the recording could be analysed for comparison with tokens of "like" in other recordings). Such data are almost always multivariate, and, more importantly not assumed to have a simple Gaussian distribution. It its therefore appropriate to use more complex pdf models to calculate likelihood ratios for these data.

Calculating a likelihood ratio for a single data point from the offender recording is a relatively simple matter, as illustrated above, irrespective of whether single-Gaussian models or more complex models are used; however, multiple data points introduce a problem. A likelihood ratio can be sequentially calculated for each data point, but what is needed is a likelihood ratio which characterises the strength of evidence with respect to the whole of the offender's speech on the offender recording, not with respect to multiple individual portions of the recording.

Since the single-Gaussian models are unimodal and symmetrical, a sensible solution for the single-Gaussian procedure could be to take the mean value of all the data points from the offender recording (a mean vector for multivariate data) and calculate a single likelihood ratio at that value. This solution is not, however, applicable for more complex pdf models which are potentially multimodal (e.g., background model in Figure 2) and potentially skewed (e.g., suspect model in Figure 2). To illustrate the multimodal problem, imagine that you have two offender data points and each corresponds to a peak in the suspect model but that the mean of those two points corresponds to a trough between those two peaks, see Figure 3. All else being equal, the two likelihood ratios calculated for the two offender data points would be relatively high, but the single likelihood ratio calculated for the mean of the offender data points would be relatively low. In the example in Figure 3 the situation is even worse because the mean of the offender data points corresponds to a peak in the background model; each of the individual offender data points results in a likelihood ratio of 1.77 and the mean of the offender data points results in a likelihood ratio of 0.36. Clearly there is something wrong here, intuitively if each individual data point results in a relatively high likelihood ratio we would expect a likelihood ratio based on the combined data also to be relatively high, not to be substantially lower.

Rather than combining the raw data before calculating a likelihood ratio, we could instead combine the multiple likelihood ratios resulting from the multiple data points. If we were to assume that the likelihood-ratio values were not correlated with each other, we could simply multiply them together à la naïve Bayes, but given that the offender data points come from multiple sections in the same recording, and in the case of MFCCs from adjacent and overlapping sections in the recording, we expect a great deal of correlation between these likelihood-ratio values.

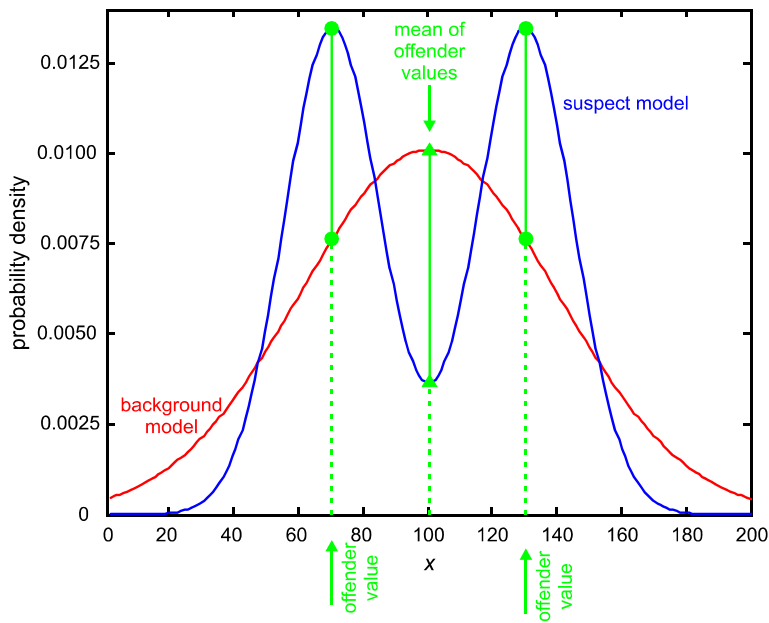

Figure 3: Example of the calculation of a likelihood-ratio values at two given offender data values (circles) and at the mean of the two values (triangles) using Gaussian-mixture suspect and background models. Data are invented for illustrative purposes.

**Calculating a score**

A typical solution to the problem above, especially in the case of MFCC GMM-UBM, is to take the mean of the logarithm of all the individual data-point-specific likelihood ratios, and call this a *score*, see Equation 2.

$$s = \frac{1}{N}\sum_{i=1}^{N}\log(LR_i) \tag{2}$$

Where *s* is the score, *i* is the index of a data point $x_i$ from the offender recording which leads to likelihood ratio $LR_i$ (calculated as in Equation 1), and the total number of data points and resulting likelihood ratios is *N*. In principle any base can be chosen for the logarithm, for algorithmic convenience calculations are often performed using natural logarithms, but for presentation of results base ten logarithms are usually preferred.

A score is log-likelihood-ratio like in that it is a number which indicates the similarity of two samples while taking into account their typicality with respect to a model of the relevant population, but the number cannot be interpreted as a log likelihood ratio. Given two scores of different values, we can say that the larger of the two provides greater support for the same-speaker hypothesis over the different-speaker hypothesis than does the smaller, or equivalently that the smaller of the two provides greater support for the different-speaker hypothesis over the same-speaker hypothesis than does the larger, but nothing else. We do not know if the absolute value of either score provides more support for the same-speaker hypothesis than for the different-speaker hypothesis or vice versa, let alone how much more support, and we do not know whether the difference between the values of the two scores indicates a large or small difference in their degree of support for one hypothesis over the other. We need another step to convert scores to interpretable likelihood ratios.

**Converting a score to a likelihood ratio using single-Gaussian models**

We could conceptualise the procedure which we use to generate a score as a procedure for converting multivariate data from multiple objects to a univariate datum describing the relationship between those objects. This score, this univariate datum, then needs to be converted to a likelihood ratio. Several procedures can be used to convert scores to likelihood ratios (see Ramos Castro[4] §6.5). Here I describe a procedure using single Gaussians as a precursor to describing logistic-regression calibration. The latter is preferred over the former because it is usually more robust, and it is the most popular procedure in the forensic-voice-comparison literature.

The procedure for converting a score to a likelihood ratio using single-Gaussian models is similar to that described above for calculating a likelihood from raw data. First we need some training data. We run a number of pairs of samples, some same-origin and some different-origin, through the initial stages of our forensic-comparison system and get scores as output. For each score we know whether it is the result of a same-origin or a different-origin comparison. The top panel in Figure 4 shows a set of same-origin and different-origin scores plotted on the abscissa. We use these scores to train a same-origin model and a different-origin model. Each of these models is a single Gaussian and, in addition, the two Gaussians are constrained such that they have the same variance (the pooled within-group variance). We can now use these models to calculate a likelihood ratio: We take a test pair and first run it through the initial stages of our forensic-comparison system to calculate a score. We then evaluate the ratio of the pdf values of the two models at the value of that score, see Equation 3.

$$LR = \frac{f(s|H_{so})}{f(s|H_{do})} = \frac{\frac{1}{\sigma\sqrt{2\pi}}e^{\frac{(s-\mu_{so})^2}{-2\sigma^2}}}{\frac{1}{\sigma\sqrt{2\pi}}e^{\frac{(s-\mu_{do})^2}{-2\sigma^2}}} = e^{\frac{(s-\mu_{so})^2-(s-\mu_{do})^2}{-2\sigma^2}} \qquad (3)$$

Where $f(s|H_{so})$ and $f(s|H_{do})$ are the probability-density-function values, from same-origin and different-origin models respectively, evaluated at a score $s$ coming from a test comparison of two samples; $\mu_{so}$ and $\mu_{do}$ are the mean values calculated from the same-origin and different origin training scores respectively; and $\sigma$ is the pooled within-group standard deviation calculated using all the training scores.

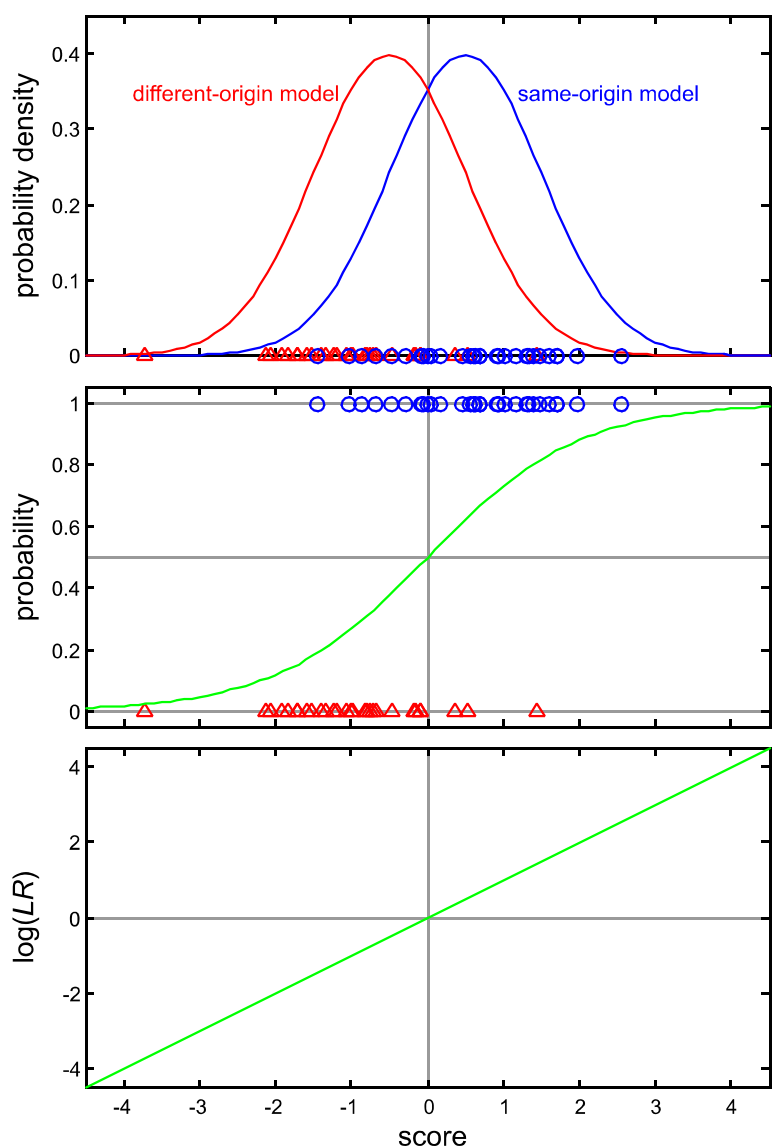

Figure 4: Gaussian probability density functions with equal variances fitted to scores (top panel), probability curve fitted to scores (middle panel), and straight line in logged odds space fitted to scores (bottom panel). Circles represent same-origin scores used for training and triangles represent different-origin scores used for training. Each panel can be mapped to the other panels, see text.

## Why equal variance?

Before describing our preferred procedure for converting from a score to a likelihood ratio, I take an aside to address the question of why we imposed an equal-variance constraint on the single Gaussians fitted to scores but not on the single Gaussians fitted to the raw data.

In terms of raw data, good properties to measure for calculating likelihood ratios are those which have relatively large between-speaker variability and relatively small within-speaker variability. Raw data consisting of measurements of properties with these sorts of distributions can potentially lead to likelihood ratios which are far from one, and therefore to high system validity. For our likelihood-ratio-calculation procedure based on raw data, we therefore want to fit one model which accounts for the hopefully larger variability of the background data (representing the distribution of the properties across a number of different speakers representative of the relevant population), and another model which accounts for the hopefully smaller variability of the suspect data (representing the distribution of the properties from multiple samples produced by a single speaker). We can see this situation reflected in Figure 1. Note that the peak of the suspect-model curve is much higher than the peak of the background-model curve because it has a smaller variance, and, because the area under the curve is fixed at 1 for each curve, a curve gets taller as it gets narrower. An inevitable consequence of the difference in the variance between the suspect model and the background model is that there will be a range of data values for which the pdf value of the suspect model is higher than the pdf value of the background model, but outside this range in both directions the opposite will be true. In Figure 1 we see that for low data values the pdf value of the background model is greater than that of the suspect model, then at about 113 Hz the relationship is reversed and the pdf value of the suspect model is greater than that of the background model, then at about 168 Hz the relative relationship is reversed again. Above the latter point the calculated likelihood ratio would lend greater support to the different-origin hypothesis than to the same-origin hypothesis, and as the distance from this point gets greater so does the degree of support for the different-origin hypothesis over the same-origin hypothesis. This is as it should be: Assuming that our models are good models of the distribution of the data from the suspect and the relevant population, the higher values are more probable if they come not from the suspect but from some other speaker in the population. Such a speaker may be atypical in that they are out on the tail of the modelled distribution for the population, but it is speakers with relatively atypical values which lead to the larger variance in the background model in the first place.

Now, turning to the single-Gaussian models fitted to the scores. Relative score values indicate the modelled degree of similarity between a suspect and offender sample given their typicality with respect to a model of the relevant population. Lower-valued scores indicate greater relative support for the different-origin hypothesis over the same-origin hypothesis and higher valued scores indicate greater relative support for the same-origin hypothesis over the different-origin hypothesis. Locally this may not always be true, in reality it may be the case that a particular lower valued score should actually have been higher than some of its nearest higher neighbours, but globally there should not be a region of low-value scores that give greater support to the different-origin hypothesis than the same-origin hypothesis, then a region of intermediate-value scores that give greater support to the same-origin hypothesis than the different-origin hypothesis, then a region of high-value scores that give greater support to the different-origin hypothesis than the same-origin hypothesis. Unlike for the suspect- and background-curves modelling the raw data, there should therefore not be a second crossing of the same-origin and different-origin curves. It would be inappropriate to apply a model which allowed for a second crossing point, at least within the range of any score which could conceivably be presented for calibration. The way to ensure that there is no second crossing point when using a model consisting of two single Gaussians is to constrain the same-origin and different-origin models such that they have the same variance.

**Mapping between single Gaussians and logistic regression**

Our preferred procedure for converting from a score to a likelihood ratio is logistic regression calibration. As will be explained in a later section, logistic regression is more robust in that it is not so dependent on the assumptions of equal-variance Gaussian distributions as is the single-Gaussians procedure. Logistic regression is a standard statistical procedure, descriptions of which can be found in numerous textbooks and tutorial papers, including references 29, 30, 31, 32, 33, 34.

Given the equal-variance constraint on the single-Gaussians model, we can map directly from that model to a logistic regression model. The middle panel in Figure 4 shows the modelled probability of the same-origin hypothesis given a score, which can be calculated directly from the Gaussians in the top panel using Equation 4 (a form of Bayes' Theorem).

$$p(H_{so}|s) = \frac{f(s|H_{so}) \times p(H_{so})}{f(s|H_{so}) \times p(H_{so}) + f(s|H_{do}) \times p(H_{do})} \quad (4)$$

Where $p(H_{so}|s)$ is the modelled posterior probability of the same-origin hypotheses given a test score $s$. If we specify equal priors, $p(H_{so}) = p(H_{do})$, then Equation 4 simplifies to Equation 5 (this specification is discussed later in this section).

$$p(H_{so}|s) = \frac{f(s|H_{so})}{f(s|H_{so}) + f(s|H_{do})} \quad (5)$$

The sigmoidal curve in the middle panel of Figure 4 could also be modelled using logistic regression. For logistic-regression training we would code the same-origin and different-origin training scores as 1 and 0 respectively. The code indicates the probability of the score coming from the same-origin category, see Figure 4 middle panel. The $p(H_{so}|s)$ curve in the middle panel of Figure 4 would then be trained on these data. Logistic-regression calculations are actually performed in a logged odds space using a maximum-likelihood technique (see works on logistic regression cited above for details), and we can map from the probability space of the middle panel of Figure 4 to the logged odds space of the bottom panel using Equation 6.

$$\log\left(\frac{p(H_{so}|s)}{p(H_{do}|s)}\right) = \log\left(\frac{p(H_{so}|s)}{1 - p(H_{so}|s)}\right) \quad (6)$$

Because we are specifying equal priors, the prior odds are 1, and the logarithm of the likelihood ratio is the same as the logarithm of the posterior odds, see Equation 7 (7a is the odds form of Bayes' Theorem).

$$\frac{p(H_{so}|s)}{p(H_{do}|s)} = \frac{f(s|H_{so})}{f(s|H_{do})} \times \frac{p(H_{so})}{p(H_{do})} \quad (7a)$$

$$p(H_{so}) = p(H_{do}) \tag{7b}$$

$$\Rightarrow \log\left(\frac{p(H_{so}|s)}{p(H_{do}|s)}\right) = \log\left(\frac{f(s|H_{so})}{f(s|H_{do})}\right) = \log(LR) \tag{7c}$$

Note that to map between the Gaussians in the top panel of Figure 4 and the logged odds space in the bottom panel of Figure 4 we do not need to go through the probability stage of the middle panel, see Equation 6, and could instead directly use the logarithm of the ratio of the pdfs, see Equation 7c right-most equality.

Setting equal priors for the logistic-regression calculation is essential for us to be able to equate the outcome of logistic regression with the logarithm of the likelihood ratio, and therefore for us to be able to use this procedure to convert from a score to a likelihood ratio. Note that there is nothing unusual about setting the priors equal since when generative models are used to calculate a likelihood ratio (e.g., Equation 1 and Figures 1 and 2) each model is a pdf subsuming an area (or in multivariate cases volume or hypervolume) of 1, the two models have equal weight. Thus for the purpose of calculating a likelihood ratio using generative models one can think of the priors as being implicitly equal. This would be the case for all generative models. In explaining the parallel discriminative procedure using logistic regression we are simply making this explicit, because in training discriminative procedures we actually have to explicitly specify the priors.

Setting equal priors for the procedure for converting from scores to likelihood ratios should not, however, be confused with setting equal priors for the same-origin and different-origin hypotheses to be combined with the likelihood ratio to arrive at posterior odds for the same-origin versus different-origin hypotheses in the case at trial. The two sets of priors are entirely independent and used for different purposes, and the trier of fact is free to assign whatever priors they deem appropriate for the trial. The logistic-regression score-to-likelihood-ratio conversion procedure does not affect the trier of fact's choice of priors for the trial.

**Logistic regression is more robust**

Although one can map directly between the procedure using single Gaussians and logistic regression there are important differences between the two. The former is a generative model, it explicitly models the distribution of each of the categories (same-origin scores and different-origin scores), whereas the latter is a discriminative model, it explicitly models the shape of the boundary between the two categories (for logistic regression this is constrained to be a particular sigmoidal shape which can be moved to the left or right and can be made shallower or steeper) but does not directly model the distribution within each category. This leads to logistic regression being much more robust to violations of the equal-variance assumption. To illustrate: Imagine that we violate the equal-variance assumption by adding to the training scores in Figure 4 one more score which we know to have come from a same-origin comparison, and that this score has a very high value compared to the existing scores, say a value of 10 or more. This is a very good score given that we know that it is a same-origin score. What effect would this have on the Gaussian procedure? The same-origin curve would move to the right such that it would be further from the different-origin curve, and the variance of both curves would increase. The corresponding sigmoidal curve in the middle panel would move to the right and become shallower. In contrast, adding this extra training score would have almost no effect on the logistic-regression model. Since logistic regression is modelling the shape of the boundary between the two categories, it is least affected by training data which are far from the boundary. Note that in the middle panel of Figure 4 it can be seen that the sigmoidal curve has asymptoted to very close to 1 by the time it has reached a score of 4.5, and the new point with a score of 10 or more would therefore have very little affect on the location and slope of the curve. Logistic regression is more robust to violations of the assumptions of equal-variance and even the assumption of Gaussian distributions and is therefore preferred over the procedure which uses the equal-variance Gaussians.

Despite being a discriminative model, because logistic regression is analogous with a generative model of the distribution of the scores in each category (as illustrated in the previous sections), and the latter clearly allows for the calculation of the ratio of the values of two probability-density functions, logistic regression is also a legitimate procedure for calculating a likelihood ratio.

**Converting a score to a likelihood ratio using logistic-regression calibration**

In the logged odds space, bottom panel of Figure 4, the relationship between a score and a log likelihood ratio is linear. Once training data have been used to train the logistic-regression model, this makes conversion of test scores to log likelihood ratios very easy. The general formula for a straight line is $y = \alpha + \beta x$, where $\alpha$ is the intercept of the line with the $y$ axis at $x = 0$, and $\beta$ is the slope of the line. We train the coefficient values for $\alpha$ and $\beta$ using logistic regression and a training set of same-origin and different-origin scores, then apply the linear translation given in Equation 8 to convert new scores from test data to log likelihood ratios. The conversion from a score to a log likelihood ratio involves shifting by the amount $\alpha$ and scaling by the amount $\beta$.

$$\log(LR) = \alpha + \beta s \tag{8}$$

The plots in Figure 4 were actually generated using Gaussians whose means were one variance apart symmetrically arranged each one half a variance away from a score of zero. The equation for the line in the bottom panel is therefore $\log(LR) = 0 + 1 \times s$, i.e., $\alpha = 0$ and $\beta = 1$, i.e., the system is already perfectly calibrated and no conversion is necessary – the scores themselves are interpretable as likelihood ratios.

We now imagine simple scenarios in which the scores are not perfectly calibrated and conversion to likelihood ratios is necessary. The plots in Figure 5 were generated in the same way as those in Figure 4 except that the training data were effectively shifted one unit to the left. Both the Gaussians have shifted one unit to the left, the sigmoidal probability curve has shifted one unit to the left, and the straight line in the logged odds space has shifted one unit to the left.

Looking at the bottom panel the scores on the abscissa can be converted to log likelihood ratios by reading off the values on the ordinate corresponding to the line. The equation for the line on the bottom panel is $\log(LR) = 1 + 1 \times s$, i.e., $\alpha = 1$ and $\beta = 1$, i.e., add one unit to convert from a score to a log likelihood ratio. This is an example of calibration shifting the scores to convert them to log likelihood ratios.

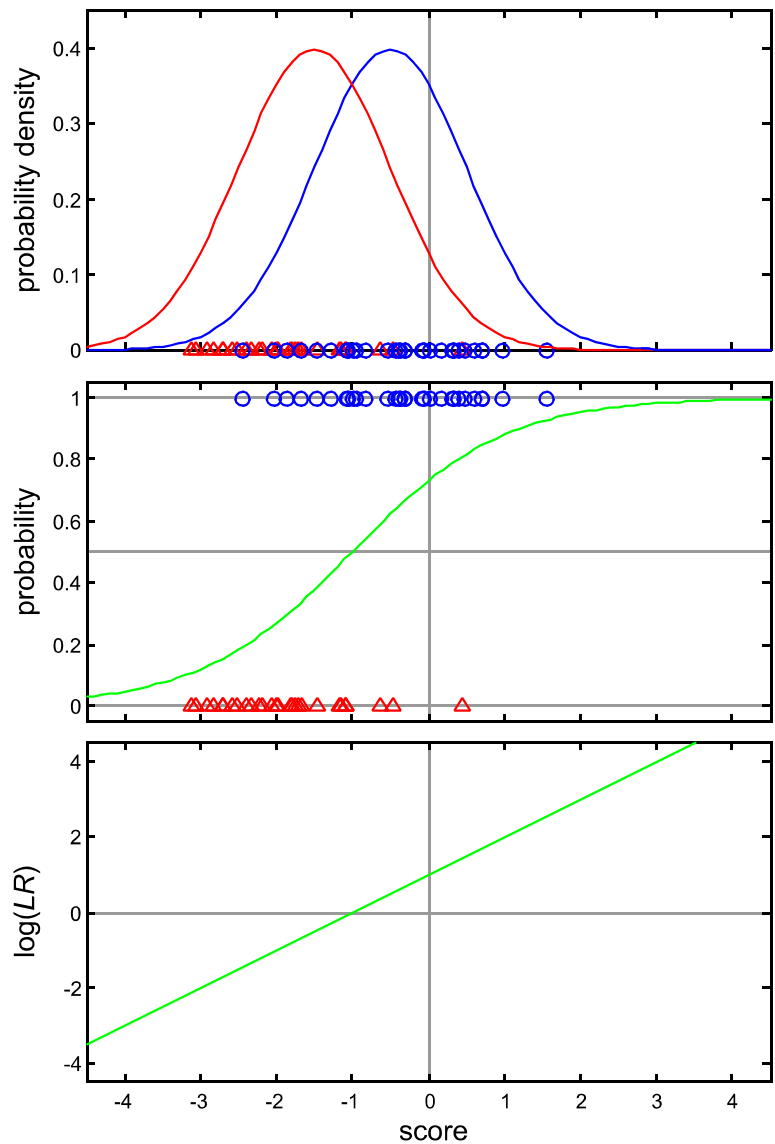

Figure 5: As Figure 4, but with all the training scores shifted one unit to the left.

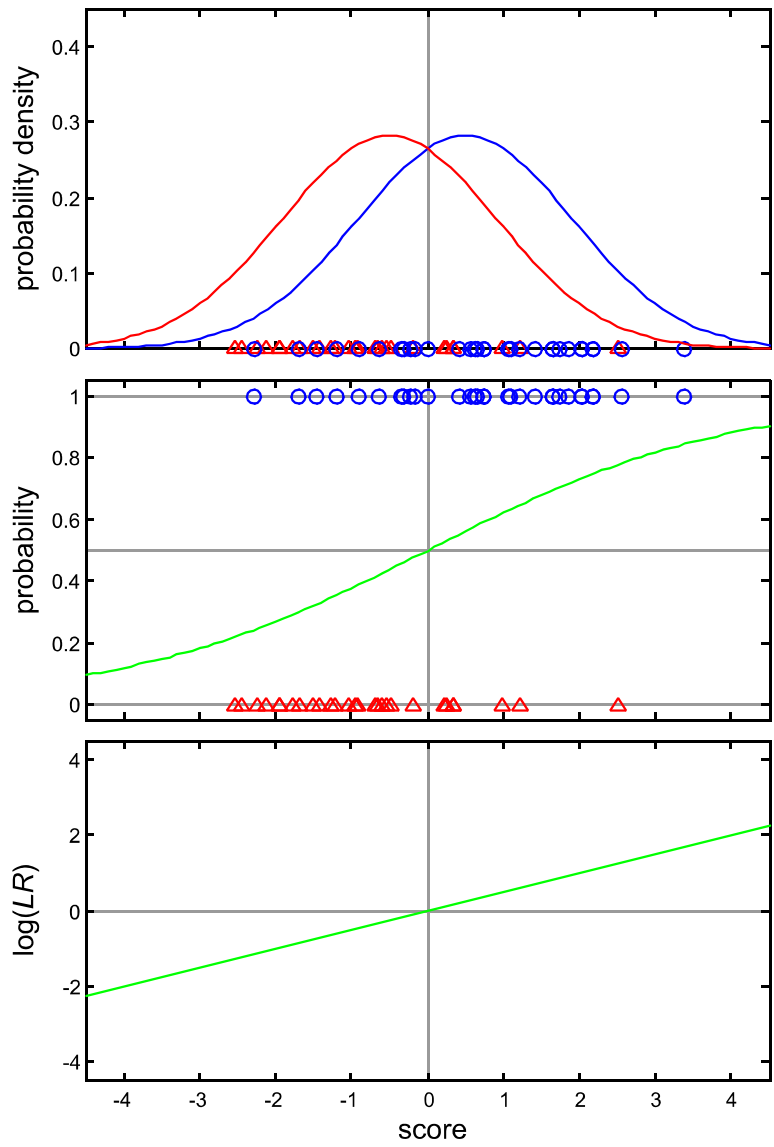

Figure 6: As Figure 4, but with the within-group variance of the training scores increased by a factor of two.

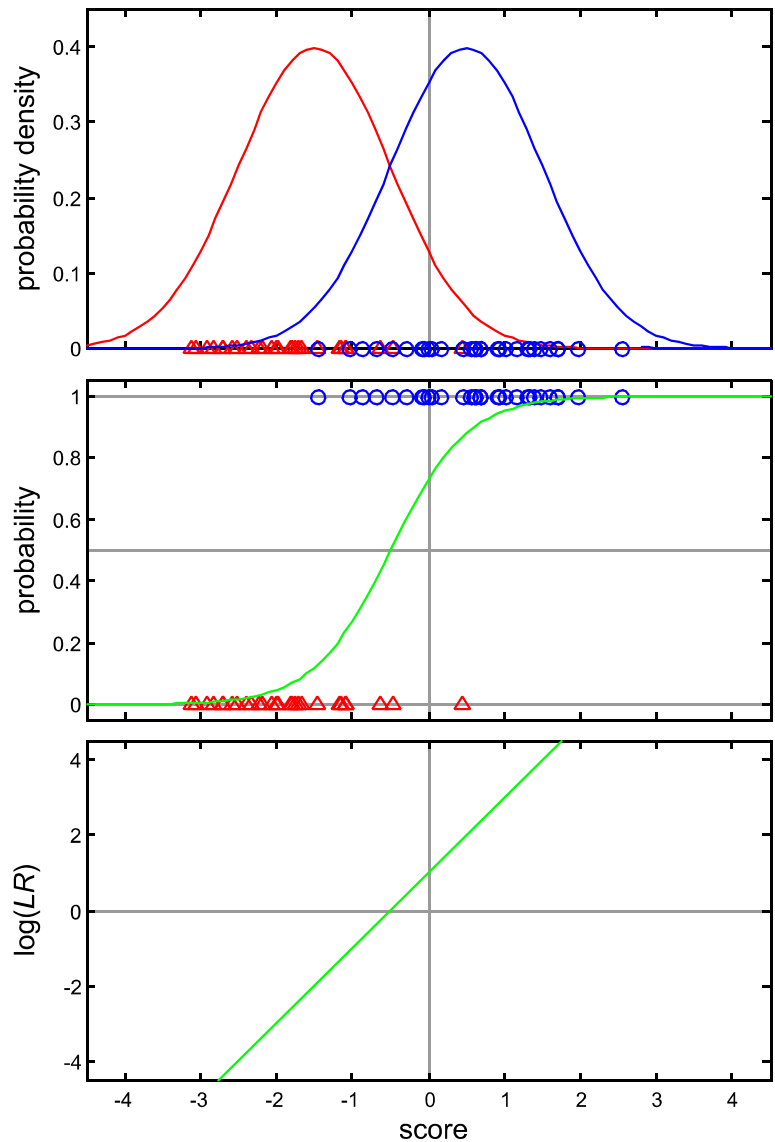

Figure 7: As Figure 4, but with all the different-origin training scores for shifted one unit to the left.

    The plots in Figure 6 were generated in the same way as those in Figure 4 except that the within-group variance of the training data was effectively increased by a factor of 2. The variance of both the Gaussians has increased, the sigmoidal probability curve has become shallower, and the slope of the straight line in the logged odds space has halved. Looking at the bottom panel the scores on the abscissa can be converted to log likelihood ratios by reading off the values on the ordinate corresponding to the line. The equation for the line is $\log(LR) = 0 + 0.5 \times s$, i.e., $\alpha = 0$ and $\beta = 0.5$, i.e., divide a score by two to convert it to a log likelihood ratio. This is an example of calibration scaling the scores to convert them to log likelihood ratios.

    Finally we consider an example which involves both shifting and scaling. The plots in Figure 7 were generated in the same way as those in Figure 4 except that the training data of the different-origin model only were effectively shifted one unit to the left. The same-origin model is unchanged relative to Figure 4. Note that the performance of the hypothetical system producing these data is better than that of the previous three artificial-data examples – the same-origin

scores are further away from the different-origin scores than they were previously, and the two Gaussians in the top panel have greater separation in terms of Mahalanobis distance (they are two variances apart compared to one, and a half, respectively in the previous examples). The sigmoidal probability curve has become steeper and moved to the left, and the slope of the straight line in the logged odds space has also become steeper and moved to the left. Looking at the bottom panel the scores on the abscissa can be converted to log likelihood ratios by reading off the values on the ordinate corresponding to the line. The equation for the line is $\log(LR) = 1 + 2 \times s$, i.e., $\alpha = 1$ and $\beta = 2$, i.e., multiply a score by two and add one to convert it to a log likelihood ratio. Note that the crossing point of the two Gaussians in the top panel is at $-0.5$, which converts to a log likelihood ratio of 0 ($-0.5 \times 2 + 1 = 0$), i.e., a likelihood ratio of 1.

Real training data need not have perfect equal-variance Gaussian distributions, and will usually result in both shifting and scaling.

**Application to scores generated in other ways**

Above, logistic-regression calibration was introduced as a solution to a problem which emerged as a result of having multiple data points from the offender sample and two relatively complex pdf models of the background and suspect data, but logistic-regression calibration has wider application and is advantageous when scores are generated in other ways. The only constraints necessary to apply logistic-regression calibration are that suitable training scores be available, that the scores indicate the similarity of pairs of objects while taking into account their typicality with respect to a relevant population, and that they be scaled as logged odds.

Morrison and Kinoshita[35] explored calibration of scores generated by Aitken and Lucy's[36,37] multivariate kernel density (MVKD) formula. In theory MVKD should produce well calibrated likelihood ratios, but this may not be the case if the modelling assumptions of MVKD are violated. Morrison and Kinoshita built an acoustic-phonetic forensic-voice-comparison system based on information extracted from the formant trajectories of tokens of the vowel /o/ in recordings of a number of speakers. They found substantial improvement in system performance when the output of MVKD was calibrated using logistic regression.

Ramos, Franco-Pedroso, and González-Rodríguez[38], and Lindh and Morrison[39] both applied logistic-regression calibration to scores generated by human listeners. A number of human listeners gave similarity judgements on a number of pairs of voice recordings. In the latter study the similarity judgments were scaled as logged odds and converted to likelihood ratios using logistic regression calibration. The results from the human-listener systems were then compared with automatic systems for which the last step had also been logistic-regression calibration.

One of the reviewers pointed out that at least some automatic speaker recognition systems and automated fingerprint identification systems (AFIS) produce scores which only take into account the similarity of pairs of samples; however, an important criterion for scores to be used in logistic-regression calibration in the forensic context is that they do take account of typicality as well as similarity. Similarity-only scores should therefore not be used. The aim of forensic comparison is to estimate the likelihood of getting the evidence given the prosecution hypothesis versus the likelihood of getting the evidence given the defence hypothesis. When working with continuously-valued data, this can be considered the similarity of the suspect and offender samples versus their typicality with respect to the relevant population. Considering only the similarity is not enough: Two samples may be very similar, but if two samples drawn randomly from the relevant population are highly likely to be equally or more similar, because both the suspect and offender sample are typical and by definition samples drawn at random are also likely to be typical, then this does not constitute strong evidence in favour of the same-origin hypothesis over the different-origin hypothesis. If, on the other hand, the two samples are very similar and atypical, then it is highly unlikely that two samples drawn randomly from the relevant population will be equally or more similar, and this combination of similarity and atypicality would constitute strong evidence in favour of the same-origin hypothesis over the different-origin hypothesis. Ignoring typicality would incorrectly lead to the supposition that the strength of evidence was equal in both cases. Since logistic-regression calibration is simply a shifting and scaling in the logged odd space, it cannot introduce an accounting of typicality into scores based solely on similarity, and the results of applying logistic-regression calibration to similarity-only scores will not produce forensically-interpretable likelihood ratios (i.e., values interpretable as the likelihood of getting the evidence given the prosecution hypothesis versus the likelihood of getting the evidence given the defence hypothesis). In applying logistic-regression calibration to the human-listeners' similarity-judgments above it was tacitly assumed that in making their judgements the listeners took perceived typicality as well as perceived similarity into account. Although it has been argued that logistic-regression is a preferred procedure because it is relatively robust, some caution should be exercised in examining the distribution of the scores in case they are clearly inappropriate even for this model.

**Logistic-regression fusion**

Conversion of scores to likelihood ratios as described above is an application of univariate logistic regression; however, logistic regression can also be applied to multivariate data. Logistic-regression fusion allows us to combine parallel sets of scores from different forensic-comparison systems, with the output being calibrated likelihood ratios. The requirement is that each system produce a score for each training comparison and for each test comparison.

The different systems could be different automatic systems (MFCC GMM-UBM is an example of an automatic system), it could be that different systems use different signal processing techniques to extract different information from the same acoustic signals, or it could be that the different systems use different modelling techniques to calculate scores. The different systems could also be different acoustic-phonetic systems each exploiting information from tokens of a different phonetic unit within the same recordings, for example, one could extract information from tokens of /aɪ/ and another from tokens of /e/.

As with logistic-regression calibration, we begin with a set of scores for training where we know which scores are the result of same-origin comparisons and which are the result of different-origin comparisons. The difference is that each system to be fused must produce a score for each comparison such that we have parallel sets of scores. These are then used to train the multivariate logistic regression model. Logistic regression automatically takes care of correlation between the sets of scores. The result of training is a set of coefficient values which provide a linear weighting of the scores from the multiple systems. These coefficient values are then used to fuse and calibrate parallel scores from test data, see Equation 9.

$$\log(LR) = \alpha + \beta_1 s_1 + \beta_2 s_2 + \ldots + \beta_n s_n \qquad (9)$$

Where $s_1, s_2, \ldots, s_n$ are the scores from the first through $n$th forensic-comparison systems, and $\beta_1, \beta_2, \ldots, \beta_n$ are the logistic-regression-coefficient weights calculated using the training set.

Figure 8 provides an artificial-data example of the training of a bivariate logistic-regression model. The left panel in Figure 8 shows the distribution of the same-origin and different-origin training scores in the two dimensions corresponding to the first set of scores, $s_1$, and the second set of scores, $s_2$. There is greater overlap between the same-origin and different-origin training scores in the first dimension than in the second dimension. Logistic-regression fusion involves calculating a weighted combination of the two dimensions which minimises the overlap between the same-origin and different-origin training scores – this is in a direction perpendicular to the $\log(LR) = 0$ line. The right panel in Figure 8 shows the resulting sigmoidal surface representing the modelled probability of same-origin scores. Running a pair of test samples through the same two forensic-comparison systems would provide a parallel set of scores which could then be converted to a log likelihood ratio using Equation 9, with, in this example, coefficient values of $\alpha = -0.93$, $\beta_1 = +0.97$, $\beta_2 = +2.32$.

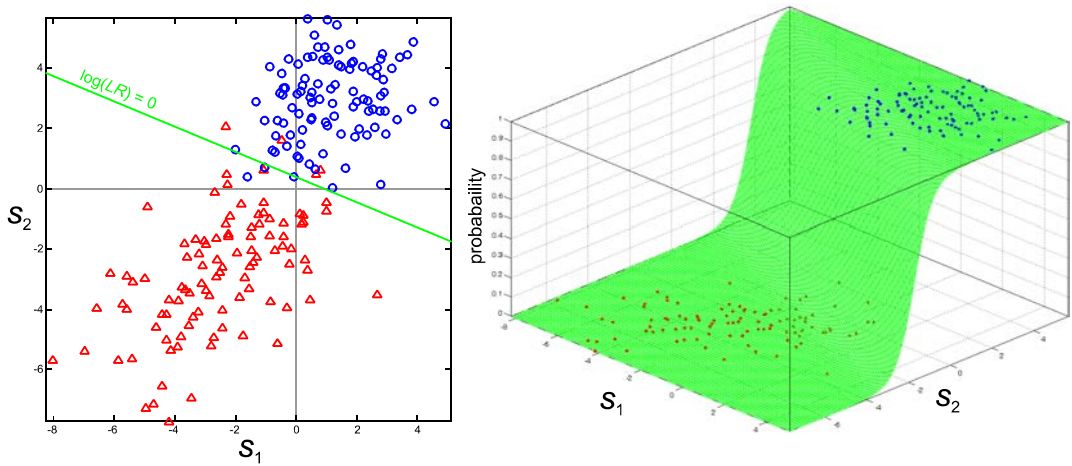

Figure 8: Parallel sets of scores [$s_1$, $s_2$] from two different forensic-comparison systems used to train a logistic-regression fusion model (left panel). Representation of the logistic-regression model in the probability space (right panel). Data are artificial and created for illustrative purposes only.

**Example of logistic-regression calibration in forensic voice comparison**

I present two examples of calibration, one applied to forensic-voice-comparison scores and the other to fingerprint/finger-mark scores.

The first example is taken from Morrison, Zhang, and Rose[8]. Recordings of 64 male speakers of Standard Chinese were analysed. Each speaker was recorded on two separate occasions separated by approximately one week. The speakers were interviewed via a landline telephone system and recorded using an analogue cassette recorder. The recordings were subsequently digitised. Tokens of several vowel phonemes were manually located and the means of each token's first three formants measured, here we only consider tokens of /a/. Scores were calculated using training data and MVKD, and these scores were used to train the coefficient weights for logistic-regression calibration. These coefficient weights were then used to convert test scores to log likelihood ratios. The test scores were also generated using MVKD. Training and test data were actually extracted from a single database, but a cross-validation procedure was adopted so that there was no overlap between the background sample used in the MVKD and the pairs of recordings compared to generate the training scores, and there was no overlap between the latter two sets of data and the pairs of recordings used to test the system. Figure 9 shows a Tippett plot for the raw scores generated by the MVKD formula and for the calibrated log likelihood ratios from one series of tests on /a/ tokens. Calibration resulted in an upwards shifting of the scores and a scaling which increased their slope. The log-likelihood-ratio cost, $C_{llr}$, decreased from 1.802 to 0.750 (the lower the $C_{llr}$ the better the system performance). System performance was very poor, but nowhere near as bad as if the uncalibrated scores had been used. (Readers unfamiliar with Tippett plots and $C_{llr}$ may wish to consult the introduction to these in Morrison[12]. In passing we note that given equal priors the optimisation objective in binomial logistic regression, the deviance statistic, is the same as $C_{llr}$.)

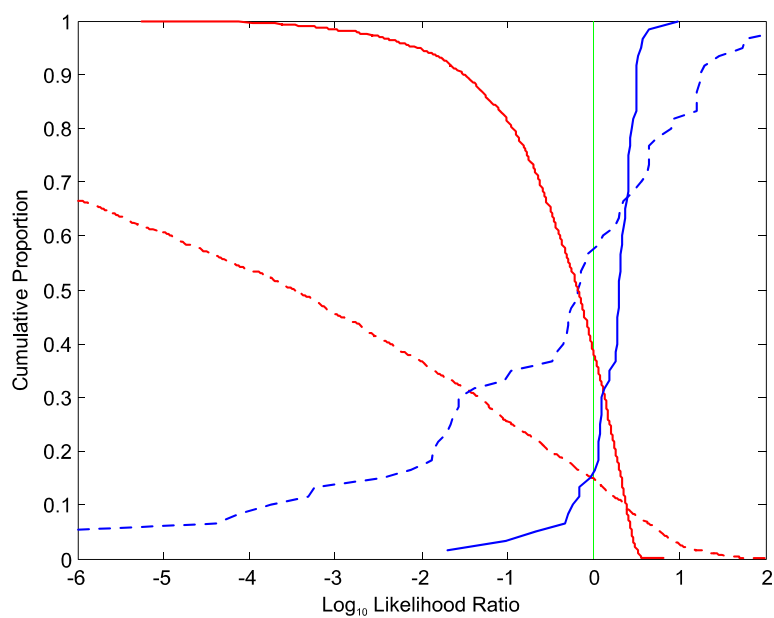

Figure 9: Tippett plot from Morrison, Zhang, and Rose[8] showing the performance of a forensic-voice-comparison system in terms of pre-calibration scores (dashed lines) and post-calibration log likelihood ratios (solid lines).

**Example of logistic-regression calibration in forensic fingerprint/finger-mark comparison**

The second example of logistic-regression calibration uses fingerprint/finger-mark data from Neumann, Evett, and Skerret[16] (my thanks to the authors for making their data available).

Neumann et al proposed a procedure for calculating likelihood ratios for fingerprint versus finger-mark comparisons; however, in the published discussion some discussants (Balding, Chacón, Holtz and Munk, Kadane, Lauritzen et al, Stern, Welling) criticised elements of the procedure for being arbitrary and questioned whether the output were indeed interpretable as likelihood ratios (it was not clear whether the functions in the numerator and denominator were in fact pdfs). Some discussants (Balding, Fieller, Lauritzen et al, Welling) also pointed out that the output of the procedure needed to be calibrated, as poorly calibrated values would be misleading to the trier of fact. Welling even fitted a logistic-regression model, and his proposed solution seems to be essentially the same as that described in the present paper. Logistic-regression calibration would not only potentially give less misleading results, but also convert the output of the Neumann et al procedure to values clearly interpretable as likelihood ratios.

The data kindly provided to me by Neumann, Evett, and Skerret consist of the "likelihood ratio" output of their system in response to pairs of test samples (numbers are rounded to five figures). I take the logarithm of each of their "likelihood ratio" output values and consider it a score. Ideally I would use a separate database to train the logistic-regression weights, but only having access to the test results I adopt a cross validation procedure. I use the results from their "large experiment". For each test finger mark there are a pair of scores available, one member of the pair resulting from a comparison of that finger mark with a fingerprint believed to be of the same origin (on the basis of decisions made by human examiners and the final outcomes of investigations/trials), and the other from a comparison of that finger mark with a fingerprint selected by an automatic fingerprint identification system as being similar but known (because of geographical/temporal information) to be of different origin. I remove a pair of scores from the database, use the remaining scores to calculate calibration weights, and then use those weights to convert the previously-removed scores to likelihood ratios. The previously-removed pair of scores is replaced in the database and the next pair removed and converted to likelihood ratios. This process is repeated until all the scores in the database have been converted to likelihood ratios. Data were divided according to how many minutiae there were in common between the finger mark and the finger prints, the number ranging from 3 to 12. Note that because the "large experiment" involved resampling to generate additional scores and I don't have access to the indices for this procedure, there may be some overlap in the logistic-regression training and testing data, something which under normal circumstances I would avoid (ideally I would exclude all resamples based on a particular mark or print from the training data when calibrating any score based on a comparison including that mark or print). The robust training procedure (see "software implementation" below) was employed with a regularization weight of 0.001.

Pre and post calibration results in terms of $C_{llr}$ are presented in Table 1, and Figure 10 shows an example Tippett plot of the performance of the 9-minutiae system pre and post calibration (which had the worst performance pre calibration but the best performance post calibration). As previously noted by Welling, the scores were biassed towards giving greater support for the same-origin hypothesis over the different-origin hypothesis, and from a full set of Tippett plots it was apparent that in general this bias increased as the number of minutiae used by the system increased. In all cases, conversion of the scores to log likelihood ratios via logistic-regression calibration removed this bias.

This can be seen as an example of a general phenomenon: If the number of data dimensions or the complexity of the model is increased and the amount of training data is not increased exponentially, the ability to obtain good estimates for model parameters decreases. Adding data dimensions or model complexity may increase the discriminative power of the output scores but they are less likely to be interpretable as likelihood ratios. Calibration operating on the score level tends to produce good results because of the use of a single dimension and a simple model and thus better estimates of model parameters (two parameters in the case of univariate binomial logistic regression).

Table 1: Performance of Neumann et al[16] fingerprint/finger-mark systems (defined according to number of minutiae exploited) pre calibration ($C_{llr}$ for scores) and post calibration ($C_{llr}$ for *LR*s).

| num. minutiae: | 3 | 4 | 5 | 6 | 7 | 8 | 9 | 10 | 11 | 12 |
|---|---|---|---|---|---|---|---|---|---|---|
| $C_{llr}$ for scores: | 0.150 | 0.112 | 0.123 | 0.138 | 0.185 | 0.196 | 0.235 | 0.212 | 0.223 | 0.174 |
| $C_{llr}$ for *LR*s: | 0.122 | 0.046 | 0.032 | 0.016 | 0.018 | 0.009 | 0.007 | 0.009 | 0.009 | 0.014 |

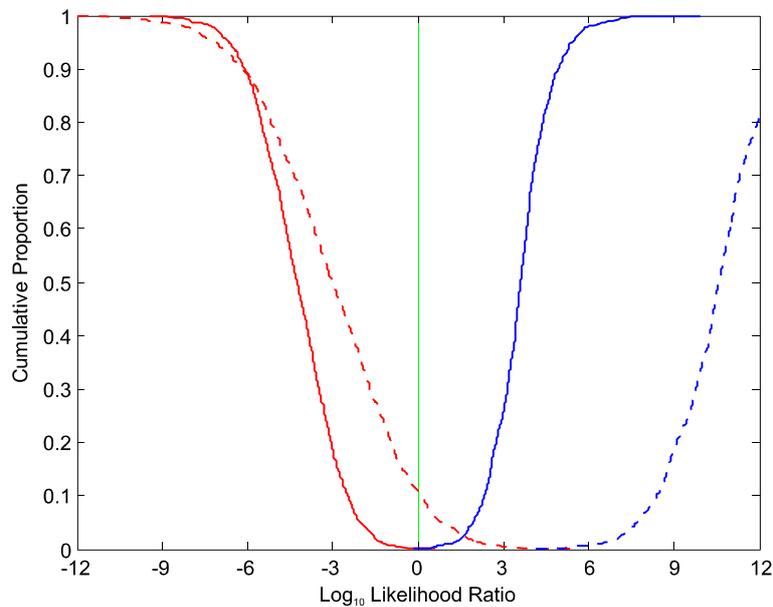

Figure 10: Tippett plot for the pre- and post-calibration results (dashed and solid lines respectively) for the Neumann et al[16] 9-minutiae fingerprint/finger-mark system.

**Example of logistic-regression fusion in forensic voice comparison**

This example of logistic-regression fusion is taken from Morrison[11]. Recordings of 27 males speakers of Australian English were analysed. Each speaker was recorded on two separate occasions separated by approximately two weeks. Recordings were digitally recorded, of studio quality, and of read speech. Tokens of the vowel phonemes /aɪ/, /eɪ/, /oʊ/, /aʊ/ and /ɔɪ/ were manually located and the trajectories of each token's first three formants measured. Third-order discrete cosine transforms (DCTs) were fitted to the trajectory of the second formant of each token. The DCT coefficient values were used as input to a GMM-UBM system which calculated scores. Five systems were constructed, one for each of the five vowel phonemes. Parallel sets of scores were calculated and used to train the coefficient weights for logistic-regression fusion. These coefficient weights were then used to convert parallel sets of test scores to log likelihood ratios. The test scores were also generated using the five GMM-UBM systems. Training and test data were actually extracted from a single database, but a cross-validation procedure was adopted so that there was no overlap between the background sample used in the GMM-UBM and the pairs of recordings compared to generate the training scores, and there was no overlap between the latter two sets of data and the pairs of recordings used to test the system. For the five systems, each using measurements from a single vowel phoneme, $C_{llr}$ based on calibrated log likelihood ratios ranged from 0.311 to 0.455, see Table 2. For the fused system, which was the result of using logistic regression to fuse the five single-vowel-phoneme systems, $C_{llr}$ improved to 0.035, an order of magnitude smaller. Figure 11 shows a comparison of the performance of the best-performing single-vowel-phoneme system (/ɔɪ/) with the system resulting from the fusion of the five single-vowel-phoneme systems.

Table 2: Performance results for best-performing individual-vowel-phonme systems in Morrison[11] and the fusion of those systems.

| system: | /aɪ/ | /eɪ/ | /oʊ/ | /aʊ/ | /ɔɪ/ | fused |
|---|---|---|---|---|---|---|
| $C_{llr}$: | 0.375 | 0.367 | 0.326 | 0.455 | 0.311 | 0.035 |

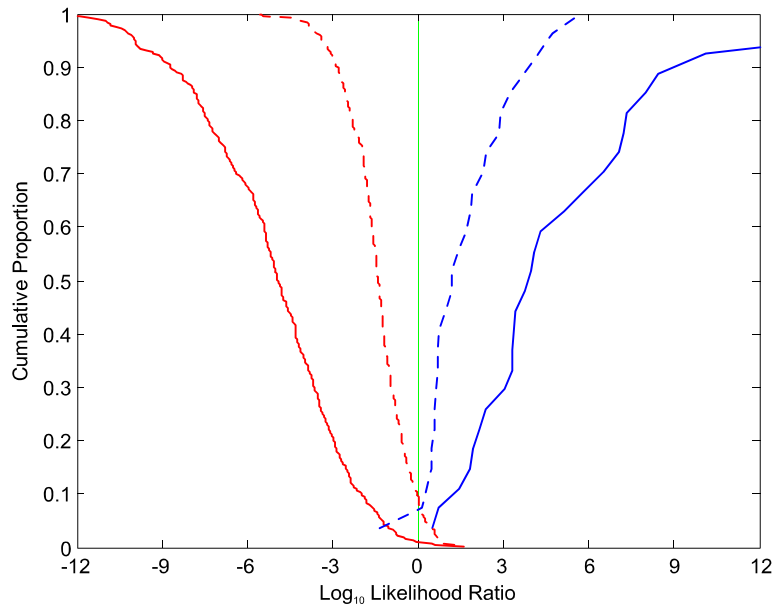

Figure 11: Tippett plot based on data from Morrison[11] showing the performance of a forensic-voice-comparison system based on a single vowel phoneme /ɔɪ/ (dashed lines) and of a system which is the fusion of five single-vowel-phoneme systems (solid lines).

**Software implementation**

Logistic regression is a standard statistical technique and as such is implemented in numerous statistical software packages. In automatic speaker recognition and forensic voice comparison a popular implementation is that in Niko Brümmer's FOCAL TOOLKIT <http://www.dsp.sun.ac.za/~nbrummer/focal/> which runs under MATLAB®. A robust version of the logistic-regression training function from the FOCAL TOOLKIT is available from <http://geoff-morrison.net/>, this is particularly useful in cases where there may be complete or near-complete separation between the same-origin training scores and the different-origin training scores. A successor to the FOCAL TOOLKIT, the BOSARIS TOOLKIT by Niko Brümmer and Edward de Villiers, is now available from <https://sites.google.com/site/bosaristoolkit/>.

**Appropriate training data**

It is important to realise that the quality of logistic-regression calibration and fusion depends on the training data. The training data should be selected from the relevant population (selection of the relevant population in forensic voice comparison is discussed in Morrison, Ochoa, & Thiruvaran[40]). Also, the recording conditions (speaking style, transmission channel, length of recording, etc.) of each of the suspect and offender samples, and any mismatch between them, should be reflected as closely as possible in each pair of samples used to generate training scores. Use of training data which do not reflect the relevant population and the conditions of the test data, i.e., the suspect and offender recordings, may lead to inaccurate log-likelihood-ratio results for the test data and for the comparison of the actual suspect and offender sample in the case at trial.